\documentclass[aps,prb,twocolumn,amsmath,amssymb,superscriptaddress,longbibliography]{revtex4-1}
\usepackage{amsmath,amssymb,bm}

\usepackage{graphicx}

\begin{document}
\title{Flat band in twisted bilayer Bravais lattices}

\date{\today}

\author{Toshikaze Kariyado}\email{kariyado.toshikaze@nims.go.jp}
\affiliation{International Center for Materials Nanoarchitectonics
(WPI-MANA), National Institute for Materials Science, Tsukuba, 305-0044,
Japan}
\affiliation{Department of Physics, Harvard University, Cambridge MA 02138, USA}
\author{Ashvin Vishwanath}
\affiliation{Department of Physics, Harvard University, Cambridge MA 02138, USA}

\begin{abstract}
 Band engineering in twisted bilayers of the five generic  two-dimensional Bravais networks is demonstrated. We first derive symmetry-based constraints on the interlayer coupling, which helps us to predict and understand the shape of the potential barrier for the electrons under the influence of the moir\'{e} structure without reference to  microscopic details. It is also pointed out that the generic constraints becomes best relevant when the typical length scale of the microscopic interlayer coupling is moderate. The concepts are numerically demonstrated in simple tight-binding models to show the band flattening due to the confinement into the potential profile fixed by the generic constraints. On the basis of the generic theory, we propose the possibility of anisotropic band flattening, in which quasi one-dimensional band dispersion is generated from relatively isotropic original band dispersion. In the strongly correlated regime, anisotropic band flattening leads to a spin-orbital model where intertwined magnetic and orbital ordering can give rise to rich physics. 
\end{abstract}

\maketitle

\section{Introduction}
Having control over electron propagation is a key to derive new functionalities of materials. Under a periodic potential, energy-momentum dispersion relation of electrons shows diverse structures: one may have a forbidden energy range as a band gap, or the effective mass can differ from the one in vacuum. In crystals, electrons feel the ionic potential, and  band engineering is achieved by chemical substitution of ions, for instance. However, the potential profile of ions depends on the properties of the ions themselves. A possible way to overcome the limitation of the band engineering set by nature is to introduce cleverly designed artificial superstructures into the system \cite{esakiSL}. 

Very recently, assembling atomically thin materials \cite{Geim:2013aa} has been recognized as an interesting and promising scheme to induce superstructures, especially after the report of  correlated electron behavior and superconductivity in twisted bilayer graphene \cite{Cao:2018ab,Cao:2018aa}. Stacking of two layers can give various superstructures as moir\'{e} patterns. When we stack two different layers, a moir\'{e} pattern is induced by lattice constant mismatch. When we stack two identical layers but with mismatched angle, again we have a corresponding moir\'{e} pattern. For the purposes of electronic structure, the moir\'{e} pattern is regarded as spatially varying interlayer coupling/interaction, which affects the electron propagation \cite{PhysRevLett.99.256802,PhysRevB.81.161405,PhysRevB.82.121407,Bistritzer12233,Koshino_2015}. Indeed, the correlated behavior in twisted bilayer graphene is associated with effective mass reduction, or band flattening, caused by the moir\'{e} pattern \cite{PhysRevB.82.121407,Bistritzer12233,PhysRevB.96.075311,Cao:2018aa,PhysRevLett.122.106405}. Generically speaking, knowing how interlayer coupling is modulated under moir\'{e} pattern \cite{PhysRevB.89.205414,PhysRevB.93.235153} is essential to predict electronic structure in the stacked systems. 

In this paper, we consider slightly twisted bilayers of identical layers of generic two-dimensional Bravais networks: oblique, rectangle, square, diamond (centered-rectangle), and triangle networks. We first show that there exist some constraints on the interlayer coupling such that it has to vanish at certain points in the moir\'{e} unit cell, depending on which part of the original Brillouin zone the relevant electrons (typically electrons at the band top/bottom) come from. The constraints are of symmetry origin, and derive solely from the properties of each layer without reference to the microscopic details of the interlayer coupling. Roughly speaking, vanishing interlayer coupling indicates that the electrons in twisted bilayer feels potential barrier there, and therefore, the derived constraints helps us to understand and predict how electrons are confined in the moir\'{e} effective potential. Second, by assuming a simple model, we classify the possible situations in terms of the typical length scale of the microscopic interlayer coupling, and identify the regime where the generic constraint becomes most relevant. The proposed concepts, the constraints and the classification of the regimes, are then numerically demonstrated in a simple tight-binding model, confirming flat bands due to  confinement by the effective potential. Generally speaking, flat bands are an excellent playground for strong correlation. We show that we can realize nested Hubbard model with a staggered orbital structure on the square lattice, or spin-valley degenerate Hubbard model in a triangular lattice model. The generic constraints further indicate a notable situation where the band flattening is highly anisotropic, i.e., we can create quasi one-dimensional bands from relatively isotropic bands in each layer. 
It is shown that with anisotropy an interesting 
 spin-orbital model with intertwined magnetic and orbital coupling is obtained in the strongly correlated limit. 

Two important ingredients in our theory are the crystalline symmetry and the position of the band top/bottom in the original Brillouin zone. Currently, the majority of  work on twisted bilayers is on hexagonal networks with electrons from K- and K'-points, like graphene, TMDC, and hBN. However, a recent data driven search for exfoliatable materials \cite{Mounet:2018aa} proposes many new candidates which are not necessarily hexagonal, which may give a suitable testbed for our theory. Another potential playground is two-dimensional carbon allotropes \cite{doi:10.1063/1.453405,PhysRevB.58.11009,C3TC30302K,C3CS60388A,C3NR04463G,doi:10.1021/acs.accounts.7b00205}. By changing connections between carbon atoms, various types of networks and a variety of positions of the band top/bottom may be achieved. Our theory will be useful in screening out interesting materials from the candidates.

\section{Constraints on the interlayer coupling}
\subsection{Electrons in moir\'{e} pattern}
Let us define the problem: we consider stacking of two identical layers taking the z-axis as a normal direction. To have angle mismatch, the upper layer is rotated by $+\phi/2$, while the lower layer is rotated by $-\phi/2$ along the z-axis. We assume that each of the layers has the full symmetry of one of the five two-dimensional Bravais lattices. Writing the unit vectors characterizing each layer as $\bm{a}_1$ and $\bm{a}_2$, the moir\'{e} pattern induced by twist has a periodicity of $\bm{L}_1$ and $\bm{L}_2$, which is explicitly written as
\begin{equation}
 \bm{L}_i=\hat{z}\times\bm{a}_i/(2\sin\frac{\phi}{2}),
\end{equation}
i.e., $\bm{L}_i$ is obtained as 90$^\circ$ rotated $\bm{a}_i$ with the scaling factor $1/(2\sin(\phi/2))$. In a small angle limit, a system is locally well approximated by an untwisted bilayer, but globally, the moir\'{e} pattern is characterized by spatial modulation in the stacking condition. For the later use, we show an explicit formula for the spatial dependence of the stacking condition. In untwisted bilayers,the stacking is described by in-plane relative shift of the two layers, $\bm{\tau}$. The origin of $\bm{\tau}$ is defined so that $\bm{\tau}=0$ corresponds to the right-on-top condition, where the upper layer is derived by the vertical shift of the lower layer. If we set $\bm{\tau}=0$ on the rotation center, the twist by $\phi$ leads to the spatial dependence of $\bm{\tau}$ following \cite{PhysRevB.89.205414}
\begin{equation}
 \bm{\tau}(\bm{r})=2\sin\frac{\phi}{2}\hat{z}\times\bm{r}.\label{eq:taudef}
\end{equation}
In the following, we use Eq.~\eqref{eq:taudef} as it is, but any constant shift of $\bm{\tau}(\bm{r})\rightarrow\bm{\tau}(\bm{r})+\bm{\tau}_0$ simply leads to a change in the origin of the moir\'{e} pattern, and does not modify the physics in the small angle limit \cite{Bistritzer12233,PhysRevB.98.085435}.

\begin{figure*}[tb]
 \centering
 \includegraphics{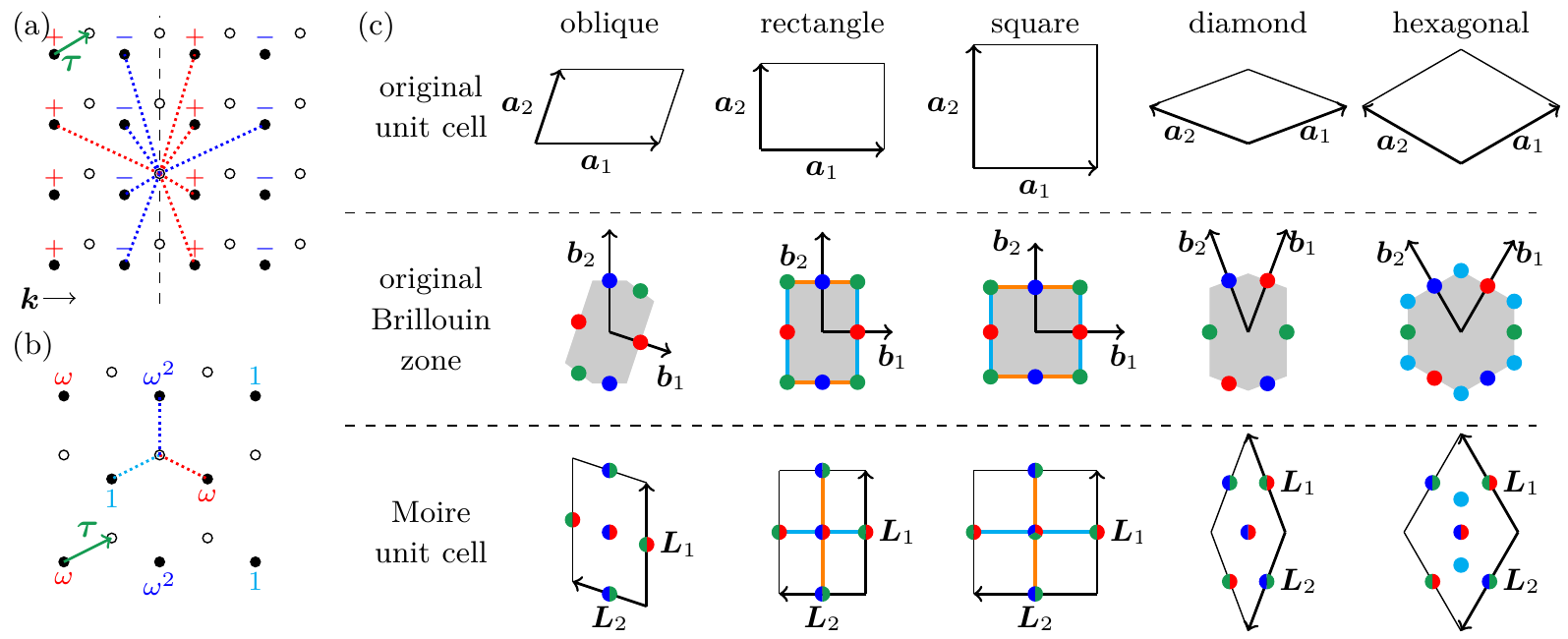}
 \caption{(a,b) Examples of the stacking adapted interference caused by intralayer momentum leading to vanishing interlayer coupling, (a) for the square lattice with $\bm{k}=\bm{b}_1/2$, and (b) for the triangular lattice with $\bm{k}=\pm(\bm{b}_1+\bm{b}_2)/3$ (K and K' points). (c) Table of the destructive interference manifolds. When the base momentum matches the colored dots or lines in the original Brillouin zones (second row), interlayer coupling vanishes on the dots or lines with the same color in the moir\'{e} unit cells (third row).}\label{fig:main}
\end{figure*}
This global modulation in the stacking condition affects the electronic structure, that is, physical properties can be tuned as a function of the twist angle. The effective model for electrons under the influence of the moir\'{e} pattern is generically written as \cite{PhysRevLett.99.256802,PhysRevB.81.161405}
\begin{equation}
 H_{\text{eff}}=
  \begin{pmatrix}
   H_+(-i\bm{\nabla}) & V(\bm{r}) \\
   V^\dagger(\bm{r}) & H_-(-i\bm{\nabla})
  \end{pmatrix},\label{eq:Heff}
\end{equation}
where the diagonal components are for the kinetic energy in each layer with the twist by $\pm\phi/2$, while the offdiagonal components are for the interlayer coupling. In principle, the diagonal part ($H_{\pm}$) can depend on $\bm{r}$ through the direct static electric potential from partner layers, but we focus on the case with dominating offdiagonal $V(\bm{r})$ (as is usually done in graphene). Since the interlayer coupling depends on the local stacking condition, $V(\bm{r})$ depends on position through the spatial dependence of $\bm{\tau}(\bm{r})$. Roughly speaking, in the small angle or the large moir\'{e} length scale limit, the kinetic energy containing spatial derivative gets less important. In the regime where the interlayer coupling dominates, it is convenient to go to the basis in which $V(\bm{r})$ part is diagonalized. Once it is brought to the diagonal, $V(\bm{r})$ looks like a potential energy \cite{PhysRevB.89.205414,PhysRevLett.121.026402}. Therefore, knowing the properties of $V(\bm{r})$ is essential to derive the electronic structure in twisted bilayers.

As we have noted, a system with small twist is locally well approximated by untwisted bilayers, and therefore, investigating properties of the interlayer coupling in the untwisted cases gives useful information. In the untwisted bilayer, no superstructure is formed and the unit cell is unchanged. This means that the translation symmetry of the unstacked lattice is kept intact, and therefore, the hamiltonian can be assigned for each momentum $\bm{k}$ in the original Brillouin zone as \cite{PhysRevB.95.245401}
\begin{equation}
 H_{\bm{k}}=
  \begin{pmatrix}
   H_0(\bm{k}) & V_{\bm{k}}(\bm{\tau}) \\
   V_{\bm{k}}(\bm{\tau}) & H_0(\bm{k})
  \end{pmatrix},
\end{equation}
where $H_0(\bm{k})$ is the hamiltonian for each layer, and the interlayer coupling $V_{\bm{k}}(\bm{\tau})$ is now a function of $\bm{k}$ and $\bm{\tau}$. To relate $V_{\bm{k}}(\bm{\tau})$ with $V(\bm{r})$, we focus on the situation that the electronic state in each layer is approximated as 
\begin{equation}
 H_{\text{single}}=E_0+\frac{1}{2}\alpha_{ij}\hat{p}_i\hat{p}_j,\quad
  \hat{\bm{p}}=-i\bm{\nabla}-\bm{k}_0.\label{eq:Hsingle}
\end{equation}
That is, we require each layer to have quadratic band dispersion characterized by an inverse mass tensor $\alpha_{ij}$ around the energy $E_0$ and the momentum $\bm{k}_0$. In the following we call $\bm{k}_0$ the \textit{base momentum}. In the small angle limit, the moir\'{e} Brillouin zone gets folded and smaller, and it follows that the energy scale of each band also gets smaller. Then, it is natural to expect that the bands closest to $E_0$ are mostly contributed by the state with $\bm{k}\sim\bm{k}_0$ in each layer, and in such a case, 
\begin{equation}
 V(\bm{r})\sim V_{\bm{k}_0}(\bm{\tau}(\bm{r})) \label{eq:VandV}
\end{equation}
should be a good approximation. Note that the important thing is to force the relevant states to have $\bm{k}\sim\bm{k}_0$ as the energy scale is reduced, and the quadratic band dispersion itself is not a requirement for Eq.~\eqref{eq:VandV} being a good approximation. Since $V_{\bm{k}}(\bm{\tau})$ is periodic in $\bm{\tau}$ with the period $\bm{a}_i$, while $V(\bm{r})$ should be periodic in $\bm{r}$ with the period $\bm{L}_i$, it is convenient to express Eq.~\eqref{eq:taudef} using $\bm{\tau}(\bm{r})=\tau_1(\bm{r})\bm{a}_1+\tau_2(\bm{r})\bm{a}_2$ and $\bm{r}=r_1\bm{L}_1+r_2\bm{L}_2$. It reduces to 
\begin{equation}
 \tau_i(\bm{r})=-r_i,
\end{equation}
which indicates all possible $\bm{\tau}$ is contained in a single moir\'{e} unit cell spanned by $(r_1,r_2)\in [0,1]\otimes[0,1]$. 

\subsection{Symmetry Based Constraints}
Now, it is worth noting that the symmetry gives constraints on $V_{\bm{k}}(\bm{\tau})$. More specifically, for the specific pairs of $(\bm{k},\bm{\tau})$, $V_{\bm{k}}(\bm{\tau})$ is guaranteed to vanish. This is an interference effect where  symmetry forces exact cancellation, and in the previous work on the untwisted cases \cite{PhysRevB.95.245401}, possible sets of $\bm{k}$ leading to $V_{\bm{k}}(\bm{\tau})=0$ are named the stacking ($=\bm{\tau}$) adapted interference manifold for $\bm{\tau}$. For instance, let us consider a single orbital tight-binding model on the fully symmetric square lattice, which has the reflection plane on the line of sites and also in between the line of sites in each layer. If we stack the two layers (without twist) so that two different types of the reflection plane are on top of each other, the microscopic interlayer hopping integrals should obey the reflection symmetry. Then, the interlayer matrix element between plane wave state with momentum $(\pi,0)$ (the lattice constant is taken as 1) on each layer vanishes because of the cancellation by the Bloch phases [see Fig.~\ref{fig:main}(a)]. The rotation axis instead of the reflection plane can also play a pivotal role in canceling the matrix element as Fig.~\ref{fig:main}(b) describes the example in the triangular lattice with momentum at the K- or K'-point. Detailed discussions and a table for the all possible pairs of $(\bm{k},\bm{\tau})$ in the five two-dimensional Bravais lattices are given in Ref.~\onlinecite{PhysRevB.95.245401}. Note that $V_{\bm{k}}(\bm{\tau})=0$ is derived using generic Bloch states (without degeneracy), and not limited to tight-binding models.

Through Eq.~\eqref{eq:VandV}, the constraints derived for the untwisted cases impose the constraints on the twisted cases as well. If the base momentum has a special value, i.e., if $\bm{k}_0$ is the one contained in the table of the stacking-adapted interference manifold, $V(\bm{r})$ has to be zero at some point in the moir\'{e} unit cell, since a single moir\'{e} cell contains all possible $\bm{\tau}$. We define manifolds (the sets of lines or points) specified by $\{\bm{r}|V_{\bm{k}_0}(\bm{\tau}(\bm{r}))=0\}$ as the \textit{destructive interference manifold} (DIM) for a base momentum $\bm{k}_0$. Since $V(\bm{r})$ works as a potential, the lines or points on which $V(\bm{r})=0$ behave as potential barriers in the effective model, which means that the shapes of the potential barrier can be fixed by symmetry without reference to the microscopic interlayer couplings when the base momentum takes a special value. Figure~\ref{fig:main}(c) summarizes the all DIMs for each of the five Bravais networks. The colored manifold in the moir\'{e} unit cell corresponds to the DIM for the base momentum with the same color in the original Brillouin zone. The colored dots are associated with the rotation symmetry (mostly $C_2$ rotation, except $C_4$ as green dots in the square case and $C_3$ as cyan dots in the hexagonal case), while the colored lines are associated with the reflection symmetry. In the rectangular and the square lattices, every colored dot is on some colored lines, i.e., the rotation symmetries add nothing than the reflection symmetries give. However, this information will be useful if we try to extend the arguments here beyond the fully symmetric Bravais networks, or going to the all wallpaper groups where in some cases the reflection planes are removed. The complete discussion on the all wallpaper groups are an interesting future work, but generically, less symmetry will give less restriction, and the fully symmetric cases are the most interesting ones in terms of imposing constraints free from microscopic details.

\subsection{Classification of regimes by the typical length scale of the interlayer coupling}
To trap electrons into the effective potential efficiently, having high spatial contrast in $V(\bm{r})$ is crucial. The constraints by DIM are useful in this context since the cancellation of $V(\bm{r})$ on special points/lines can be a source of high contrast. While DIM itself relies only on symmetry and is generic, having high contrast requires that the spatial range of the microscopic interlayer coupling is moderate compared to the microscopic lattice scale. If the microscopic interlayer coupling is too long ranged on the scale of the microscopic lattice, an averaging effect suppresses $V(\bm{r})$ as a whole when $|\bm{k}_0|\neq 0$, while if it is too short ranged, influence of DIM originating from the interference of Bloch states get less pronounced since the wave nature is not resolved in very small length scale.

For better understanding, we analyze a simple case  that each layer is described by a single orbital tight-binding model. Here, we write the interlayer hopping as $f(\bm{r})$ where $\bm{r}$ is a vector connecting two sites on the different layers projected on the plane of the layer. Using the gauge where the lower and the upper layer wave functions are written as $\psi_-(\bm{R})\sim e^{i\bm{k}\cdot\bm{R}}$ and $\psi_+(\bm{R})\sim e^{i\bm{k}\cdot(\bm{R}+\bm{\tau})}$, respectively with $\bm{R}$ being lattice points, $V_{\bm{k}}(\bm{r})$ becomes \cite{PhysRevB.86.155449,PhysRevB.87.205404,PhysRevB.89.205414,PhysRevB.95.115429,PhysRevX.8.031087,nonpert}
\begin{align}
  V_{\bm{k}}(\bm{\tau}) 
 = 
 \sum_{\bm{R}}e^{-i\bm{k}\cdot(\bm{R}+\bm{\tau})}f(\bm{R}+\bm{\tau})
  = \sum_{\bm{G}}e^{i\bm{G}\cdot\bm{\tau}}f_{\bm{k}+\bm{G}}.\label{eq:ff}
\end{align}
$f_{\bm{q}}$ is the Fourier component of $f(\bm{r})$ and $\bm{G}$ denotes the reciprocal vectors. 
When $f(\bm{r})$ is pretty long ranged, $f_{\bm{q}}$ decays rapidly as a function of $|\bm{q}|$. But for $\bm{k}_0$ satisfying the condition to have DIM, $|\bm{k}_0+\bm{G}|$ is always finite and Eq.~\eqref{eq:ff} tell us that $V_{\bm{k}_0}(\bm{\tau})$ is suppressed as a whole , which is not advantageous to obtaining high contrast. 
In addition, if $V_{\bm{k}_0}(\bm{\tau})$ is suppressed, the spatial dependence of the diagonal part of Eq.~\eqref{eq:Heff} may come into play. On the other hand, when $f(\bm{r})$ is very short ranged, Eq.~\eqref{eq:ff} gives us $|V_{\bm{k}}(\bm{\tau})|\sim |f(\bm{\tau})|$ with $\bm{\tau}$ limited in the Wigner-Seitz cell in which we do not see $\bm{k}$ dependence.
Then, as we have noted, at the intermediate regime where $f(\bm{r})$ is not too long ranged nor too short ranged, the generic constraints derived by symmetry become most relevant. Roughly speaking, in the intermediate regime, the typical length scale should be smaller than the original lattice constant $a$, but still, should be a sizable fraction of $a$. For graphene, the sublattice degrees of freedom are important and each layer cannot be approximated as Eq.~\eqref{eq:Hsingle}, but when it comes to the range of the interlayer hopping, it is in the ideal regime in the sense that $f(\bm{r})$ is significantly suppressed at $|\bm{r}|\sim |\bm{a}_i|$, but is sizable, say at $|\bm{r}|=|\bm{a}_i|/2$ \cite{PhysRevB.93.235153}.

\section{Numerical Demonstration}
\subsection{Band flattening}
\begin{figure}[tb]
 \centering
 \includegraphics{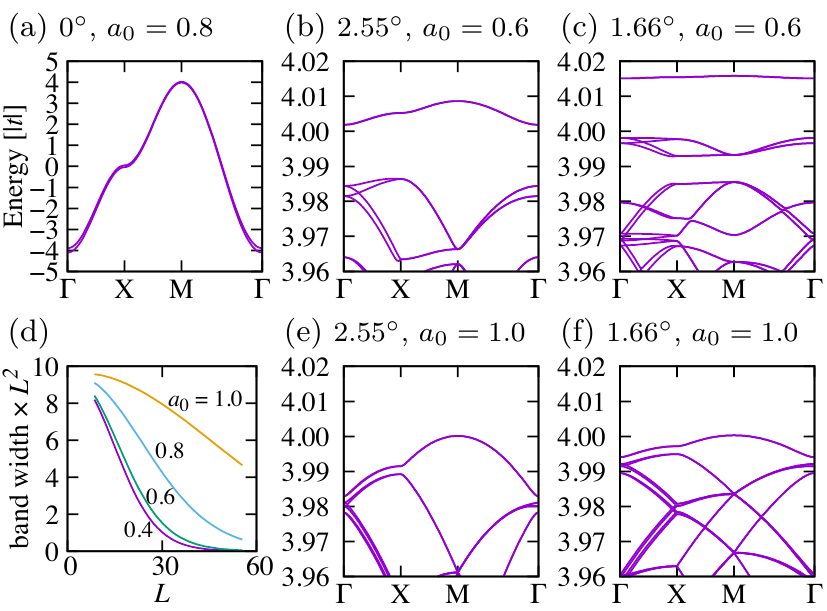}
 \caption{(a) Band structure of the untwisted bilayer square lattice model with $a_0=0.8$ for $\bm{\tau}=0$. (b,c) Band structures of the twisted case with $a_0=0.6$ at $\phi=2.55^\circ$ and $1.66^\circ$, respectively. (d) Band width of the top most band scaled by a factor of $L^2$ for several values of $a_0$. $L=30$ and $60$ correspond to $\phi=1.91^\circ$ and $0.95^\circ$, respectively. (e,f) The same as (b,c) but with $a_0=1.0$.}\label{fig:sq}
\end{figure}
From now on, we numerically demonstrate how the electrons are confined in the symmetry enforced potential barriers. The first example is the square lattice model, where the intralayer hopping is limited to the nearest neighbor hoppings of $t=-1$, while the interlayer hopping $f(\bm{r})$ is assumed to be 
$f(\bm{r})=V_0 \exp\bigl(-\frac{r^2}{a_0^2}\bigr)$
for simplicity \cite{PhysRevB.93.235153,Padhi_Phillips}. Here, the typical length scale of $f(\bm{r})$ is determined by $a_0$, and we set $V_0=|t|/20$. We focus on the state at the band top, which is located at the M-point in the original Brillouin zone. In Fig.~\ref{fig:main}(c), the M-point for the square lattice is on both of the orange and the cyan lines, and correspondingly, the DIM forms square grid lines, from which we expect that the electrons are bind to square lattice. 

Figure~\ref{fig:sq} summarizes the numerical results. We can clearly see that the top most two bands, which are nearly degenerated and difficult to be isolated in the shown energy scale, pop up and get flattened as the angle is reduced in Figs.~\ref{fig:sq}(b) and \ref{fig:sq}(c). 
We will return to this double degeneracy soon.
It is confirmed that the band flattening gets weaker when the typical length scale of the microscopic interlayer hopping gets larger in Figs.~\ref{fig:sq}(e) and \ref{fig:sq}(f). The band width of the top most band is plotted as a function of the moir\'{e} length scale $L$ in Fig.~\ref{fig:sq}(d). Note that we originally have quadratic dispersion at the M-point, and the size of the moir\'{e} Brillouin zone scales as $1/L$, there is a trivial contribution to the band width reduction by a factor of $1/L^2$. Therefore in Fig.~\ref{fig:sq}(d), the band width is scaled by a factor of $L^2$. Still, we observe strong reduction of the scaled band width when $a_0$ is small.

In real space, the flat bands are contributed from one localized orbital per a single moir\'{e} unit cell, and the double degeneracy of the top most band comes from the fact that the minimal commensurate supercell for the square moir\'{e} pattern contains two moir\'{e} unit cells (see Appendix). The nearest neighbor pair of moir\'{e} cells are microscopically distinguishable, and affects the symmetry of the localized orbitals. Specifically, when we focus on the states from the M-point in the original Brilloun zone, the nearest neighbor pairs are occupied by orbitals with different $C_4$ rotation eigenvalues, which we refer to as an $s$-type orbital and a $d_{xy}$-type orbital. The different symmetry of the orbitals suppresses the nearest neighbor hopping in moir\'{e} cells, and the leading term will be the next nearest neighbor hopping between the same species of the orbitals, which results in a nested network of the two decoupled square lattices (one from the $s$-orbitals and the other from the $d_{xy}$-orbitals), whose lattice constant is $\sqrt{2}$ of the one of the moir\'{e} pattern. Even for incommensurate structures where
such a simple $s$- and $d_{xy}$-orbitals picture does not hold, the
effect of nearest neighbor site decoupling is still apparent in the numerically obtained band structure in the small angle limit.
In the case that the screening from the substrate is strong, electron-electron interaction will be short ranged, and we expect that the system is modeled as a nested square lattice Hubbard model.

\begin{figure}[h]
 \centering
 \includegraphics{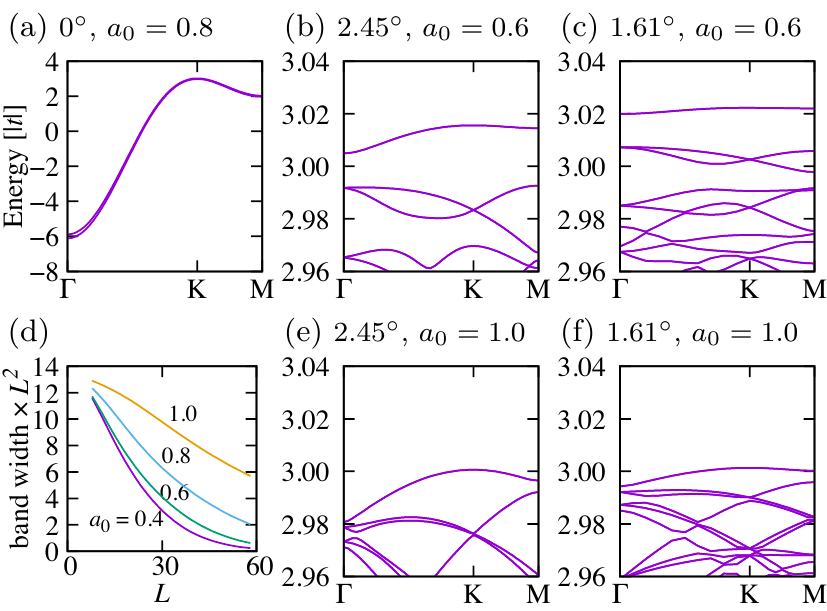}
 \caption{(a) Band structure of the untwisted bilayer triangular lattice model with $a_0=0.8$ for $\bm{\tau}=0$. (b,c) Band structures of the twisted case with $a_0=0.6$ at $\phi=2.45^\circ$ and $1.61^\circ$, respectively. (d) Band width of the top most band scaled by a factor of $L^2$ for several values of $a_0$. (e,f) The same as (b,c) but with $a_0=1.0$.}\label{fig:tri}
\end{figure}

In the square lattice case, the DIM forms grid lines. We can also test the case where the DIM forms arrays of dots in the bilayer of the nearest neighbor tight-binding model on the triangular lattice. If we set the nearest neighbor hopping negative, the band top is at the K- and K'-points, and the DIM forms honeycomb lattice, from which we expect that the electrons are bind in the triangular lattice. 
Figure~\ref{fig:tri} summarizes the numerical results for this case, which shows a clear sign of the band flattening for the moderate $a_0$. Note that the top most bands in Figs.~\ref{fig:tri}(b), \ref{fig:tri}(c), \ref{fig:tri}(e), and \ref{fig:tri}(f) are doubly degenerate, which stems from the degeneracy between the K- and K'-points in the original band structure. This will give us a Hubbard model on the triangular lattice with extra degrees of freedom from the valley degeneracy in the strongly correlated regime \cite{PhysRevLett.121.026402,PhysRevLett.122.086402,hBN}.

\subsection{Anisotropic band flattening}
\begin{figure}[tb]
 \centering
 \includegraphics{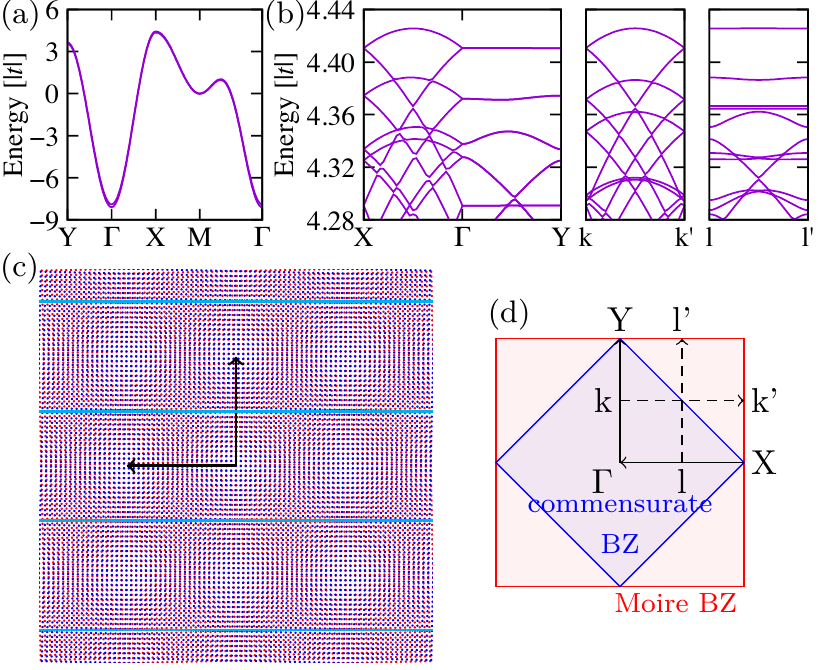}
 \caption{(a) Band structure of the untwisted bilayer square lattice model with strong n.n.n. hopping. For $\delta<0$, the band top is at the X point ($\bm{k}=\bm{b}_1/2$). Here, $a_0=0.8$ and $\bm{\tau}=0$ is adapted. (b) Band structure with the twist angle $\phi=2.55^\circ$ and $a_0=0.6$ along the path shown in (d). (c) Schematic picture of the destructive interference lines in the moir\'{e} structure for the states from the original X point.}\label{fig:X}
\end{figure}
An interesting application of Fig.~\ref{fig:main}(c) is that we find situations that the band flattening becomes anisotropic. For the square and the rectangular cases, if the Base momentum is on the Brillouin zone edge but not at the zone corner, the DIM runs in one specific direction, which means that the electrons feel quasi one-dimensional potential. To demonstrate this, we consider the bilayer of a single orbital tight-binding model on the square lattice with large next nearest neighbor hoppings, for which hamiltonian is written as
\begin{multline}
 H_0(\bm{k})=2(t+\delta)\cos k_x+2(t-\delta)\cos k_y\\
 +2h(\cos(k_x+k_y)+\cos(k_x-k_y)).
\end{multline}
By setting $t=-1$, $h=-1$, and $\delta=-0.1$, the band top locates at the X-point [see Fig.~\ref{fig:X}(a)]. The finite $\delta$ is introduced simply to make it easier to see the effects by distinguishing the X- and the Y-points. For the interlayer hopping integrals, we again adapt the same gaussian form as the previous example. The DIM for the case where the base momentum is on the X-point is schematically written in Fig.~\ref{fig:X}(c), showing the quasi one-dimensional nature. Figure~\ref{fig:X}(b) shows the obtained band structure under twist along the path in the moir\'{e} Brillouin zone specified in Fig.~\ref{fig:X}(d). We clearly see that the band is flattened in the $y$-direction, while it remains highly dispersive in the $x$-direction, despite the original band structure has only moderate anisotropy around the X-point. 

\begin{figure}[h]
 \centering
 \includegraphics{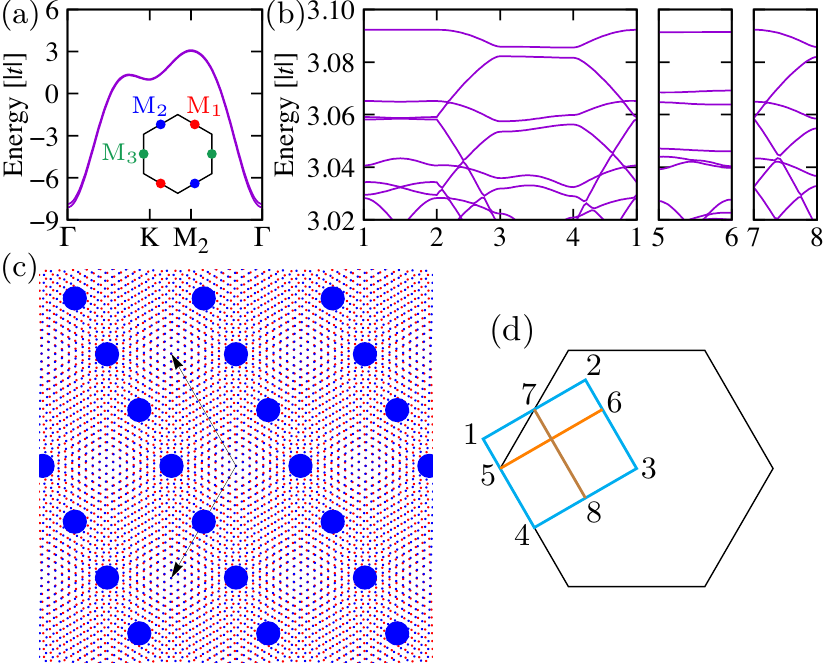}
 \caption{(a) Band structure of the untwisted bilayer triangular lattice model with n.n. and n.n.n. hopping. For $\delta<0$, the band top is at the M$_2$ point ($\bm{k}=\bm{b}_2/2$). Here, $a_0=0.8$ and $\bm{\tau}=0$ is adapted. (b) Band structure with the twist angle $\phi=2.61^\circ$ and $a_0=0.6$. The numbers on the horizontal axis correspond to the points labeled by the same numbers in (d). (c) Schematic picture of the destructive interference lines in the moir\'{e} structure for the states from the original M$_2$ point.}\label{fig:M}
\end{figure}

A similar anisotropic band flattening is possible also in a triangular lattice model. According to Fig.~\ref{fig:main}(c), if the base momentum is located on one of the M-points, say $\bm{k}_0=\bm{b}_2/2$, the DIM looks like Fig.~\ref{fig:M}(c). Since the aspect ratio of the arrays of dots (representing the locations where the interlayer coupling vanishes) is not unity, the anisotropic band flattening is expected to follow. To have the base momentum located on the M-point, in each layer we assume hoppings up to the second nearest neighbor ones. Then, each of the layers is described by
\begin{multline}
 H_0(\bm{k})=2(t+\delta)\cos k_1+2(t-\delta)\cos k_2+2t\cos(k_1+k_2)\\
 +2h(\cos(2k_1+k_2)+\cos(k_1+2k_2)+\cos(k_1-k_2)),
\end{multline}
where $k_i=\bm{k}\cdot\bm{a}_i$. For $t=-1$, $h=t/3$, and $\delta=0.1$, the band top is located at the M$_2$-point, which can be used as the base momentum [see Fig.~\ref{fig:M}(a)]. Note that $\delta$ is introduced just for a clear demonstration by isolating one of the three M-points. Note also that the band dispersion is isotropic around each of the M-point when $\delta=0$. The gaussian type interlayer hopping is adapted as the other examples. The resultant band structure is shown in Figure~\ref{fig:M}(b), where the path depicted in Fig.~\ref{fig:M}(d) is used. We see that along 1--2, 3--4, and 5--6 direction, the bands are relatively flat, while along 1--4, 2--3, and 7--8 direction, the bands are relatively dispersive. The contrast between the flat and the dispersive parts is most evident in the top most band.

\subsection{State at the $\Gamma$-point}
We have noted that for very short ranged $f(\bm{r})$, the position of the base momentum gets less important and the strong spatial modulation of $V(\bm{r})$ can be induced without help of DIM. This statement can be numerically checked by looking at the band bottom at the $\Gamma$-point in the single orbital tight-binding model of the square and the triangular lattices. 
It is confirmed that as $a_0$ gets reduced, the band flattening gradually becomes prominent (see Appendix).

\section{Discussion}
\subsection{Materials}
The two key ingredients in our demonstration are the crystalline symmetry and the position of the base momentum $\bm{k}_0$. Thus far the most of the studies on the twisted bilayers focused on the hexagonal network with $\bm{k}_0$ at the K- and K'-point, like graphene, TMDC, and hBN. 
As for the symmetry, among the recently predicted candidates of atomically thin layer materials \cite{Mounet:2018aa}, some of them have square networks, such as square TMD \cite{PhysRevB.89.125423} or Cu$_{\text{2}}$MX$_{\text{4}}$ (M=Mo,W, X=S,Se) \cite{PhysRevB.92.085427}.

As for the position of $\bm{k}_0$, MoS and WS are recently proposed in theory and predicted to have a hexagonal type network with its valence top at the M-point \cite{MoSWS}. These materials breaks the $C_2$-rotation symmetry, which is required for DIM associated with the M-point, but if the symmetry breaking affects the interlayer coupling only perturbatively, the anisotropy band flattening can potentially work. Or, $\gamma$-graphyne, which has a triangular network and has been considered in theory for a long time, is predicted to have the valence top and the conduction bottom at the M-point \cite{doi:10.1063/1.453405,PhysRevB.58.11009}. This comes with all the required symmetries, but we have to investigate carefully how the internal structure within a single unit cell quantitatively affects the interlayer coupling. 

\subsection{Perspectives}
Among the demonstrated results, the anisotropic band flattening is most intriguing and can be of interests in many aspects. In our demonstration of the anisotropic band flattening, the valley degeneracy (X/Y in the square lattice or three M-points in the triangular lattice) was deliberately lifted to make the effect easy to be observed. Without that deliberate lifting, we have valley degeneracy, but different valleys result in difference in anisotropy. For convenience, let us define an ``easy direction'' for each valley as the direction in which the band contributed from the corresponding valley is dispersive after the anisotropic band flattening. Then, our result is restated as the easy direction depends on the valley. This property might be used to build a valley filter or a valley splitter for valleytronics \cite{Rycerz:2007vn}. 

On the other hand, with the lifting, or focusing on the rectangular (diamond) lattice instead of the square (triangular) lattice, what we have in the moir\'{e} lattice is a set of the quasi-1D channels. In the case that the screening by the substrate is weak, the electron-electron interaction can be long ranged to couple the neighboring channels. Depending on the parameter, each 1D channel can be nearly free, and then, the entire system can be served as a coupled wire model. Here, the channel distance is tunable by the twist angle. Also, the unit cell is large in the small angle cases, which implies that a laboratory accessible magnetic field can amount to the significant fraction of flux quantum per a unit cell. 

\begin{figure}[tb]
 \centering
 \includegraphics{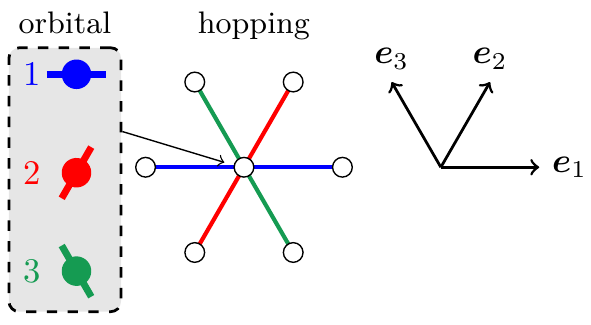}
 \caption{Schematic picture of the three orbital model.}\label{fig:threeorb}
\end{figure}
Finally, we mention on some novel physics expected in the correlated regime with flattened bands. Let us focus on the triangular lattice with base momentum at the M-point, where the valley dependent anisotropic band flattening is expected. Under the assumption that the intervalley scattering is weak, which usually applies in the small twist angle limit, three valleys are independent degrees of freedom, and we can build a three orbital model. Because of the anisotropic band flattening, the three orbitals have different easy directions. The simplest model capturing these feature is a three orbital model on the triangular (moir\'{e}) lattice, where each orbital only hops between the nearest neighbor sites in corresponding easy direction. Figure~\ref{fig:threeorb} schematically shows the hopping structure, where the orbital with specific color has only hopping on the bond with the same color. Note that three orbital model in triangular lattice is unique in the sense that there is no three (or higher) dimensional irreducible representation in any 2D point groups. Neglecting the intervalley scattering is crucial here, since in the symmetry view point, any coupling should lift the three fold degeneracy from the valley degrees of freedom.

As for the correlation, if we assume that the screening by the substrate is strong, the Coulomb interaction is modeled as onsite intra- and inter-orbital Hubbard repulsion, $U$ and $U'$. If we further neglect the Hund coupling and the pair hopping terms, the Hamiltonian becomes, 
 \begin{multline}
  H=t\sum_{\bm{r}\sigma}\sum_{\mu=1}^3
  c^\dagger_{\bm{r}+\bm{e}_\mu,\mu\sigma}c_{\bm{r},\mu\sigma} + \text{ h.c. }\\
  +U\sum_{\bm{r}}\sum_{\mu=1}^3 n_{\bm{r},\mu\uparrow}
  n_{\bm{r},\mu\downarrow}
  +U'\sum_{\bm{r},\sigma\sigma'}\sum_{\mu<\mu'}
  n_{\bm{r},\mu\sigma}
  n_{\bm{r},\mu'\sigma'},
 \end{multline}
where $\bm{e}_\mu$ is defined as in Fig.~\ref{fig:threeorb}. In the limit of $t\ll U$ and $t\ll U'$ with one electron (or hole) per a moir\'{e} unit cell, which corresponds to 1/6 filling since we have three orbitals per a site, we can derive a spin-orbital model following the standard $t/U$-expansion. The local basis for the spin-orbital model is $|\sigma\rangle\otimes|\tau\rangle$ ($\sigma=\uparrow,\downarrow$, $\tau=1,2,3$). Up to the leading order in $t/U$ and $t/U'$, we obtain a variant of the Kugel-Khomskii model \cite{Kugel1982,PhysRevB.56.R14243,PhysRevB.80.064413} as 
 \begin{multline}
  H_{\text{eff}}=J\sum_{\bm{r}}\sum_{\mu=1}^3
  \Bigl(\bm{S}_{\bm{r}+\bm{e}_\mu}\cdot\bm{S}_{\bm{r}}-\frac{1}{4}\Bigr)\tilde{\tau}^{(\mu)}_{\bm{r}+\bm{e}_\mu}\tilde{\tau}^{(\mu)}_{\bm{r}}\\
  -V\sum_{\bm{r}}\sum_{\mu\neq\mu'}
  \bigl(\tilde{\tau}^{(\mu')}_{\bm{r}+\bm{e}_\mu}\tilde{\tau}^{(\mu)}_{\bm{r}}+\tilde{\tau}^{(\mu')}_{\bm{r}-\bm{e}_{\mu}}\tilde{\tau}^{(\mu)}_{\bm{r}}\bigr), \label{eq:KK}
 \end{multline}
with $J=4t^2/U$ and $V=t^2/U'$. $\tilde{\tau}^{(\mu)}$ is a $3\times 3$ matrix acting on the orbital space, whose components are $(\tilde{\tau}^{(\mu)})_{ij}=\delta_{ij}\delta_{i\mu}$. Because of the mismatch of the easy direction between the different orbitals, there appears no orbital flip or exchange at this order of the $t/U$-expansion, and the orbital dynamics is generated only by higher order terms in the expansion, for instance as a ring exchange term. This means that the orbital degrees of freedom is classical in the Hamiltonian Eq.~\eqref{eq:KK}, which allows us to derive some eigenstates analytically. For instance, in the case that $V$ dominates, a ground state will be macroscopically degenerate and with maximum short range antiferro-orbital correlation. It is an intriguing future problem to explore the full phase diagram of Eq.~\eqref{eq:KK} and to consider quantum corrections given by the ring exchange terms. 

\section{Summary}
To summarize, we derive generic constraints on the interlayer coupling that works as the effective potential for electrons in moir\'{e} patterns for bilayers of the all five Bravais networks. The constraints are symmetry based and fixed without reference to the microscopic details, which helps us to understand and predict how electrons are trapped in an effective potential and form flat bands. Also, it is pointed out that the generic constraints become particularly important when the microscopic interlayer hoppings has a moderate length scale, compared to the microscopic lattice scale. The power of the constraints is demonstrated in simple tight-binding model by showing band flattening explicitly. From the generic constraints, we find an interesting situation where the band flattening is highly anisotropic. When the interlayer hoppings are very short ranged on the scale of the microscopic lattice, the generic constraints are less important, and it is possible to have band flattening without constraints. Yet, having the symmetry based constraints will boost material searches, and help to build a stable playground for the band flattening. In the presence of interactions this can provide a promising route to realizing a range of models from quasi-one dimensional coupled wire  to square lattice Hubbard models with staggered orbitals and the Kugel-Khomskii spin orbital models, with a high degree of tunability and control.

\begin{acknowledgments}
 The work was partially supported by JSPS
 KAKENHI Grant Numbers JP17K14358 and JP18H01162 (T.K.). A.V. was supported by NSF DMR-1411343.
\end{acknowledgments}

\appendix

\section{Appendix}

\subsection{Commensurate unit cell for the square lattice}
\begin{figure}[h]
 \centering
 \includegraphics{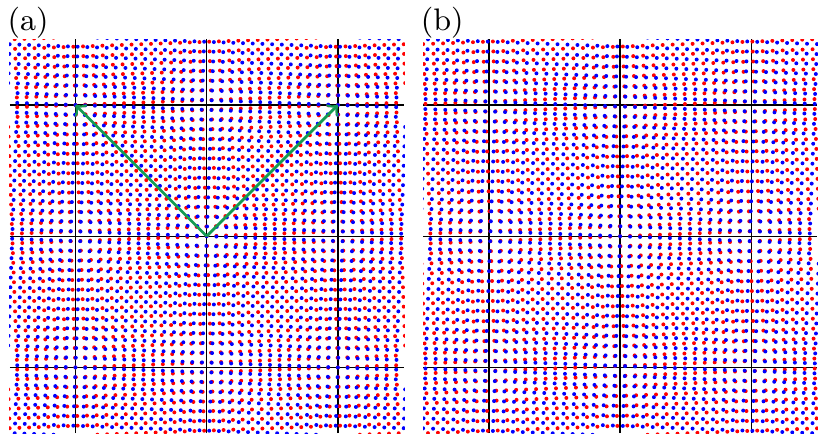}
 \caption{(a) Moir\'{e} pattern at the angle satisfying the minimal commensurate condition. The moir\'{e} period is indicated by grid lines, while the commensurate unit vectors are written as arrows. At the crossing of the grid lines, the stacking condition is $\bm{\tau}=0$, but some crossings have sites right on top, while the others do not, i.e., two regions with the same $\bm{\tau}=0$ are microscopically distinguishable. (b) Moir\'{e} pattern at the incommensurate angle to compare with (a).}\label{fig:commensurate}
\end{figure}

An easy and straightforward way to derive band structures with moir\'{e} patterns is to use a commensurate unit cell, which is large and strictly periodic. For the square lattice case, the unit vectors for the minimal commensurate cell with twist $\tilde{\bm{L}}_i$ are derived using the original unit vectors $\bm{a}_i$ as
\begin{align}
 \tilde{\bm{L}}_1=(s+1)\bm{a}_1+s\bm{a}_2,\\
 \tilde{\bm{L}}_2=-s\bm{a}_1+(s+1)\bm{a}_2
\end{align}
with $s$ an integer and $2\sin(\phi/2)=1/\sqrt{s^2+s+1/2}$. But, $|\tilde{\bm{L}}_i|=\sqrt{2}|\bm{L}_i|$ and the area spanned by $\tilde{\bm{L}}_1$ and $\tilde{\bm{L}}_2$ is twice as large as the one spanned by $\bm{L}_1$ and $\bm{L}_2$. Typical commensurate and incommensurate moir\'{e} patterns are shown in Fig.~\ref{fig:commensurate}, where the moir\'{e} period is indicated by grid lines. For the commensurate case, in the moir\'{e} cell at the center, we have dots on the crossing of the grid lines. On the other hand, in the neighboring moir\'{e} cells, there are no dots on the crossing of the grid lines. That is, there are two types of microscopically distinguishable moir\'{e} cells. 

\subsection{State at the $\Gamma$-point}
\begin{figure}[h]
 \centering
 \includegraphics{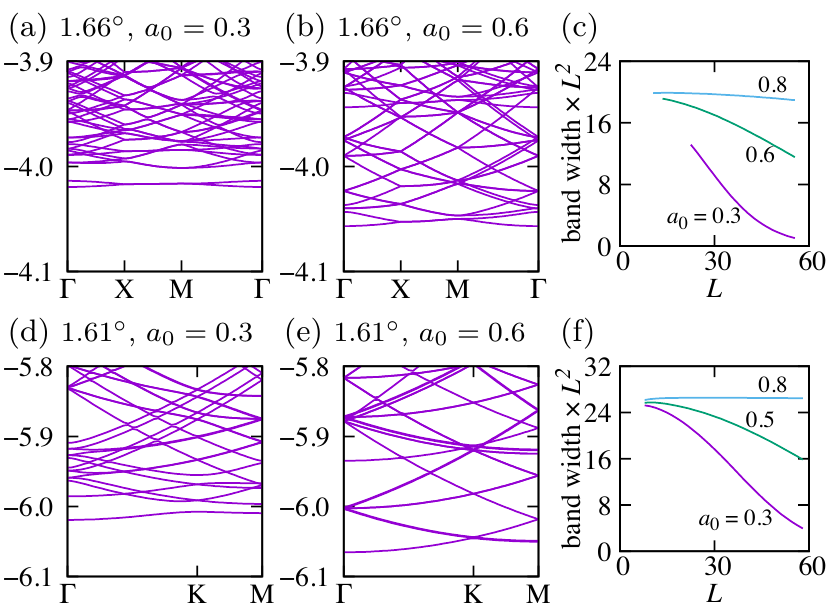}
 \caption{Band structures of the twisted bilayer square lattice model with (a) $a_0=0.3$ and $a_0=0.6$. The lower most bands are featured to see the case that the base momentum is located at the $\Gamma$-point. (c) Scaled band width of the lower most band. (d-f) The same as (a-c) but for the triangular lattice model.}\label{fig:gamma}
\end{figure}

It often happens that the base momentum is located at the $\Gamma$-point. However, the $\Gamma$-point, which is the center of the Brillouin zone, is  not related to any of the DIMs as Fig.~\ref{fig:main}(c) shows. This means that the generic constraints cannot fix the shape of the effective potential barriers, and we have to have very short ranged interlayer hopping. To demonstrate the band flattening at the $\Gamma$-point, we focus on the band bottom of the nearest neighbor tight-binding models on the square and the triangular lattices. Figure~\ref{fig:gamma} summarizes the numerical results, for which the set up is the same as the one for Figs.~\ref{fig:sq} and \ref{fig:tri}. We find that the band flattening is prominent for small $a_0$. 

\end{document}